# A Swendsen-Wang update algorithm for the Symanzik improved sigma model.

Alessandra Buonanno[1]
*Istituto di Fisica dell'Università di Pisa*

Giancarlo Cella[2]
*Istituto di Fisica dell'Università di Pisa*
(June 1, 1995)

## Abstract

We study a generalization of Swendsen-Wang algorithm suited for Potts models with next-next-neighborhood interactions. Using the embedding technique proposed by Wolff we test it on the Symanzik improved bidimensional non-linear $\sigma$ model. For some long range observables we find a little slowing down exponent ($z \simeq 0.3$) that we interpret as an effect of the partial frustration of the induced spin model.

PACS numbers: 05.50.+q, 02.50.+s

# I. INTRODUCTION

The Swendsen-Wang algorithm [3] is known to be a very efficient way of generating configurations for a Monte Carlo simulation, owing to its little slowing down exponent. For the bidimensional Ising model the numerical data are consistent with $\tau \propto \xi^{\approx 0.35}$, to be compared with $\tau \propto \xi^{\approx 2.1}$ of the conventional, local algorithms (see for example [4–8]).

However, the original formulation in the framework of the Potts spin model can't be easily generalized to other statistical systems, for example lattice gauge theories, in spite of many efforts to do it [9,10].

Wolff [11–13] has shown that it is possible to incorporate the Swendsen-Wang dynamics in a $O(N)$ invariant, multicomponent statistical system, embedding Ising variables in the continuous degrees of freedom. This method proved to be extraordinarily efficient, with an almost complete absence of critical slowing-down $\tau \propto \xi^{\lesssim 0.1}$.

Motivated by a concrete application to the non linear, Symanzik improved $O(3)$ bidimensional sigma model, we study a simple generalization of the algorithm.

# II. THE ALGORITHM

The basic idea of the Swendsen-Wang procedure is to introduce some auxiliary degrees of freedom in the model one wants to simulate. We apply the algorithm to the sigma model defined in our case by the Symanzik tree level improved action

$$S = \beta \sum_{n,\mu} \left[ -\frac{4}{3} \varphi^a(n) \varphi^a(n+n_\mu) + \frac{1}{12} \varphi^a(n) \varphi^a(n+2\,n_\mu) \right] \tag{1}$$

and by the constraint $\varphi^a(n)\varphi^a(n) = 1$. We fix an arbitrary unit vector $r$ and parameterize the field as

$$\varphi(n) = \varphi_\perp(n) + |\varphi(n) \cdot r| \sigma_n \quad \text{with} \quad \sigma_n = \frac{\varphi(n) \cdot r}{|\varphi(n) \cdot r|} = \pm 1. \tag{2}$$

At fixed $\varphi_\perp$ and $|\varphi \cdot r|$ the system is equivalent to a disomogeneous Ising model

$$S = \sum_{n,n'} J[n,n'] \delta_{\sigma_n, \sigma_{n'}} \tag{3}$$

with a ferromagnetic nearest neighbor coupling and an antiferromagnetic third neighbor one

$$J[n,n'] = 2\beta |\varphi(n) \cdot r| |\varphi(n') \cdot r| \sum_\mu \left( -\frac{4}{3} \delta_{n',n+n_\mu} + \frac{1}{12} \delta_{n',n+2n_\mu} \right). \tag{4}$$

The partition function can be written, neglecting an irrelevant multiplicative constant, as a product of terms

$$Z = \sum_{\{\sigma\}} \prod_{n,n'} (1 + k(n,n') \delta_{\sigma_n, \sigma_{n'}}) \quad \text{with} \quad k(n,n') = e^{-J[n,n']} - 1. \tag{5}$$

In each of them we introduce a new degree of freedom $l_{n,n'}$, which can get values in $\{1,0\}$, so that in the general case we can write ($\bar{\delta}_{a,b} = 1 - \delta_{a,b}$)

$$1 + k(n, n')\delta_{\sigma_n,\sigma_{n'}} = \sum_{l_{n,n'}=0,1} \Big(W_{1,1}(n, n')\delta_{\sigma_n,\sigma_{n'}}\delta_{l_{n,n'},1} + W_{0,0}(n, n')\bar{\delta}_{\sigma_n,\sigma_{n'}}\delta_{l_{n,n'},0}$$
$$+ W_{1,0}\delta_{\sigma_n,\sigma_{n'}}\delta_{l_{n,n'},0} + W_{0,1}\bar{\delta}_{\sigma_n,\sigma_{n'}}\delta_{l_{n,n'},1}\Big) \qquad (6)$$

with the conditions $W_{1,1} + W_{1,0} = 1 + k$ and $W_{0,0} + W_{0,1} = 1$. As the $W_{i,j}$ constants are proportional to probabilities they must be non–negative.

The variables we have added are in a one–to–one correspondence with the interactions of the model, and we sum over all configurations $\{l, \sigma\}$. We obtain a new partition function which describes the joint dynamics of all the degrees of freedom.

We start considering the evolution of the set $\{\sigma\}$ at fixed $\{l\}$. If we make the choice

$$W_{1,0}(n, n') = W_{0,0}(n, n') \qquad (7)$$

the interaction between the sites $n$ and $n'$ becomes irrelevant if $l_{n,n'} = 0$. This means that the spin system decomposes in a set of independent clusters $C_i$, each of them made of all the lattice sites which can be joined by a chain of $l_{n,n'} = 1$ interactions.

Inside each $C_i$ the dynamics is described by an Ising–like effective action

$$S_{\text{cluster}} = -\sum_{\sigma_n,\sigma_{n'} \in C_i} \log\left(\frac{W_{1,1}(n, n')}{W_{1,1}(n, n') - k(n, n')}\right)\delta_{\sigma_n,\sigma_{n'}} \qquad (8)$$

which can be simplified by imposing another condition. If $k > 0$ (i.e. if the interaction is ferromagnetic) we can choose $W_{1,1} = k$, obtaining a model in which the two spins must be aligned in order not to pay an infinite action tribute. If the action is antiferromagnetic the analogous choice is $W_{1,1} = 0$: in this case the two spins must be necessarily unaligned.

For a fixed $\{\sigma\}$ configuration the probability distribution for $l_{n,n'}$ depends only on the two spins $\sigma_n$ and $\sigma_{n'}$. For a ferromagnetic interaction the relevant term is (cfr. equation (6))

$$\delta_{\sigma,\sigma'}[k\delta_{l,1} + \delta_{l,0}] + \bar{\delta}_{\sigma,\sigma'}\delta_{l,0}. \qquad (9)$$

It follows that if the two spins are unaligned $l$ must be set to zero. In the case of alignment there is on the contrary an "activation" probability proportional to $k/(1+k)$. If the interaction is antiferromagnetic we obtain

$$\bar{\delta}_{\sigma,\sigma'}[(1+k)\delta_{l,0} - k\delta_{l,1}] + (1+k)\delta_{\sigma,\sigma'}\delta_{l,0}. \qquad (10)$$

In this case if the spins are aligned it follows necessarily $l = 0$, in the other case we have $l = 1$ with probability $-k$.

From these considerations it follows that after the generation of the $\{l\}$ set inside each cluster the spins automatically satisfy the constraint imposed by the equation (8), and that the only possible moves are the flippings of a cluster as a whole.

In conclusion we can sum up the procedure as follows. After choosing a random direction $r$ we set the $l$ values with the appropriate probabilities. Next we construct the clusters, and flip each of them with some assigned probability.

In absence of Symanzik improvement there are only ferromagnetic couplings, so each cluster is composed of aligned spins. In our case it is possible for two or more clusters of this type with opposite spin orientation to be joined by an antiferromagnetic active $l$.

This fact has two interesting consequences. First of all at $\beta = \infty$ our algorithm is no more ergodic, as one can easily construct field configurations that are left unchanged by it, apart for a trivial global flip. To see this consider three spins in the sites $n$, $n + n_\mu$ and $n + 2n_\mu$. If $\sigma_n \neq \sigma_{n+2n_\mu}$ the antiferromagnetic bound $l_{n,n+2n_\mu}$ is surely activated, and the same must be true for one of the two ferromagnetic ones $l_{n,n+n_\mu}$ or $l_{n+n_\mu,n+2n_\mu}$, so that all the spins belong to the same cluster. If $\sigma_n = \sigma_{n+n_\mu} = \sigma_{n+2n_\mu}$ both $l_{n,n+n_\mu}$ and $l_{n+n_\mu,n+2n_\mu}$ are activated and the three spins are connected again. Only if $\sigma_n \neq \sigma_{n+n_\mu} \neq \sigma_{n+2n_\mu}$ there is a probability that the spins can be changed independently, but this cannot occur for a sufficiently smooth configuration. So at $\beta = \infty$ the algorithm can change only high wavelength modes, while if the field configuration is smooth all spins are connected in one unique cluster, and the only possible update is a global parity. If $\beta$ is big but finite we expect the formation of a large cluster which connects nearly all the sites, and then a reduced decorrelation. We emphasize that this is not the case for the non improved model, where also at $\beta = \infty$ the only stable configuration is that in which all the spins are aligned.

Another point is that the mean size of the cluster is no more connected with the susceptibility, as is the case without improvement where a Fortuin-Kasteleyn representation exists [14].

### III. PERFORMANCES

In order to test the efficiency of the generalized algorithm we have measured the integrated autocorrelation time for the observables [15]

$$\mathsf{M}^2 = \sum_{n,m} \phi^a(n)\phi^a(m) \tag{11}$$

$$\mathsf{F} = \frac{1}{2}\left(\sum_{n,m} e^{2\pi i(n_1-m_1)/L}\phi^a(n)\phi^a(m)\right) + \frac{1}{2}\left(\sum_{n,m} e^{2\pi i(n_2-m_2)/L}\phi^a(n)\phi^a(m)\right) \tag{12}$$

$$\mathsf{E}_1 = \frac{1}{2}\sum_{n,\mu} \phi^a(n)\phi^a(n+n_\mu). \tag{13}$$

From the mean values of $\mathsf{M}^2$ and $\mathsf{F}$ one can easily evaluate the two point function at the two smaller momenta available on a finite lattice

$$\chi = V^{-1} < \mathsf{M}^2 > = \tilde{G}(p)\Big|_{|p|=0} \tag{14}$$

$$\chi' = V^{-1} < \mathsf{F} > = \tilde{G}(p)\Big|_{|p|=2\pi/L}. \tag{15}$$

These are "long distance" dynamical quantities (in particular $\chi$ is the susceptibility) from which it is possible to calculate

$$\xi = \left[\frac{(\chi/\chi') - 1}{4\sin^2(\pi/L)}\right]^{\frac{1}{2}} \tag{16}$$

which is a possible definition of correlation length in a finite volume. On the other side the mean value of $\mathsf{E}_1$ is connected to the short distance dynamics

$$E = V^{-1} < \mathsf{E}_1 >= G(n)|_{|n|=1} . \tag{17}$$

We have chosen the single cluster update scheme proposed by Wolff [11–13], and we have measured also the size of the flipped cluster $N_c$. In our parameter range the ratio $\xi/L$ is always less than 0.5, and the asymptotic scaling regime is not yet reached. For example we observe at best a 15% discrepancy between our measured correlation length and the exact value predicted by the Bethe ansatz [16].

We report in Table I the integrated autocorrelation time for $\chi,\chi',\mathsf{E}_1$ and $N_c$, extracted from a series of $10^6$ consecutive cluster updates. In Table II we list the analogous results obtained using the over heat bath algorithm [17].

We have calculated the integrated correlations applying the self–consistent method proposed by Madras et al. [18], and we have checked the stability of the result.

In order to evaluate the critical slowing down exponent for a given observable $O$ we try to fit our data using the standard finite size scaling ansatz

$$\tau_o = \xi(\beta,L)^{z_o} \phi_o \left[ \frac{\xi(\beta,L)}{L} \right] . \tag{18}$$

Here $\xi(\beta,L)$ is the measured correlation length defined by (16) and $\phi$ is a unknown universal function. As an example we report in Fig. 1 $\xi^{-z}\chi$ versus $\xi/L$ for all measures we have taken, using the value $z = 0.3$ which gives a reasonable result.

Our best estimate for the critical slowing down exponents are reported in Table III and IV. As one can see the cluster algorithm performs certainly better in respect of the local one. For the long range quantities $\mathsf{M}^2$ and $\mathsf{F}$ we argue that $0.2 < z < 0.4$. It is interesting to note that for the local quantity $E_1$ the results are consistent with a total elimination of slowing down. This is in some sense an intermediate situation between a local algorithm, which decorrelates short scales much better than long ones (see table IV for the over heat bath case), and the usual Swendsen Wang which reduces slowing down with the same efficiency at all scales.

In Fig. 2 we plot the ratio between the measured susceptibility and the mean size of flipped cluster versus the correlation length. As we have anticipated with our generalized algorithm $\chi$ is not proportional to $N_c$, as one can easily see. We try to interpret the plot in the following way: for $\xi < 0.2\,L$ the finite size effects are small, and we can see that the cluster size grows more rapidly than the "physical" size connected to the susceptibility. This is consistent with the discussion of the previous section about the expected behavior at large $\beta$ values. For $\xi > 0.2\,L$ volume effects prevent more effectively the cluster size than susceptibility from growing, hence $N_c/\chi$ decreases.

## IV. CONCLUSIONS

Our results show that the proposed algorithm is effective in reducing the slowing down at short and long scales. In the last case the slowing down is not completely eliminated and we can interpret this fact in two equivalent ways.

As there is not a proportionality between the cluster size and the physical scale of the model the algorithm is not forced to operate on the modes physically relevant.

From another point of view we have seen that the non optimal behavior at large $\beta$ is connected to the simultaneous presence of ferromagnetic and antiferromagnetic interactions in the effective spin model, that becomes frustrated. It is well known that in presence of frustration the reduction of slowing down is an extremely difficult task.

In our case the frustration is small, and the algorithm is in any case more efficient than a local one. We have worked out a more elaborate generalization of Swendsen Wang algorithm that could be effective in reducing the excessive growth of cluster size, and we are testing it to see if it is possible to further reduce slowing down in this model [19].

We are also extending our study to larger correlation length, in order to be sure that the dynamical exponents we have extimated are really the asymptotic ones.

## ACKNOWLEDGMENTS

We thank Prof. Giuseppe Curci and Prof. Andrea Pelissetto for clarifying conversations.

# REFERENCES

[1] Electronic address: buonanno@sunthpi1.difi.unipi.it.
[2] Electronic address: cella@sun10.difi.unipi.it.
[3] R. Swendsen and J.-S. Wang, Phys. Rev. Lett. **58**, 86 (1987).
[4] G. F. Mazenko and O. T. Valls, Phys. Rev. B **24**, 1419 (1981).
[5] C. Kalle, J. Phys. A **17**, 801 (1984).
[6] J. K. Williams, J. Phys. A **18**, 49 (1985).
[7] R. B. Pearson, J. L. Richardson, and D. Toussaint, Phys. Rev. B **31**, 4472 (1985).
[8] S. Wansleben and D. P. Landau, J. Appl. Phys. **61**, 3968 (1987).
[9] R. G. Edwards and A. D. Sokal, Phys. Rev. D **38**, 2009 (1988).
[10] F. Niedermayer, Phys. Rev. Lett. **61**, 2026 (1988).
[11] U. Wolff, Nucl. Phys. B **322**, 759 (1989).
[12] U. Wolff, Phys. Rev. Lett. **62**, 361 (1989).
[13] U. Wolff, Nucl. Phys. B **334**, 581 (1990).
[14] C. M. Fortuin and P. W. Kasteleyn, Physica **57**, 536 (1972).
[15] S. Caracciolo, R. G. Edwards, A. Pelissetto and A. D. Sokal, Nucl. Phys. B **403**, 475-541 (1993).
[16] P. Hasenfratz, M. Maggiore, and F. Niedermayer, Phys. Lett. B **245**, 522 (1990).
[17] R. Petronzio and E. Vicari, Phys. Lett. B **254**, 444 (1991).
[18] N. Madras and A. D. Sokal, J. Stat. Phys. **50**, 109 (1988).
[19] A. Buonanno and G. Cella, in preparation.

FIGURES

FIG. 1. Finite size scaling of $\tau_\chi^{int}$

FIG. 2. The ratio $N_c/\chi$ versus the measured correlation length.

TABLES

TABLE I. Results of numerical simulations for the autocorrelation times with Swendsen Wang algorithm.

| L | $\beta$ | $\tau^{int}_{\chi}$ | $\tau^{int}_{\chi'}$ | $\tau^{int}_{E_1}$ |
|---|---|---|---|---|
| 32 | 1.250 | 11.2 (5) | 22.6 (14) | 0.55 (6) |
| 32 | 1.300 | 12.1 (6) | 23.9 (16) | 0.51 (6) |
| 32 | 1.350 | 13.8 (7) | 24.0 (16) | 0.79 (1) |
| 32 | 1.400 | 14.6 (7) | 24.6 (16) | 0.50 (6) |
| 32 | 1.450 | 14.6 (7) | 23.2 (15) | 0.50 (6) |
| 32 | 1.500 | 15.9 (8) | 21.1 (13) | 0.50 (6) |
| 64 | 1.250 | 21.4 (13) | 17.8 (10) | 0.520 (6) |
| 64 | 1.300 | 14.3 (7) | 17.3 (10) | 0.500 (6) |
| 64 | 1.350 | 12.8 (3) | 20.7 (13) | 0.510 (6) |
| 64 | 1.400 | 13.4 (7) | 26.2 (18) | 0.500 (6) |
| 64 | 1.450 | 15.8 (8) | 27.8 (20) | 0.500 (6) |
| 64 | 1.500 | 16.7 (9) | 25.5 (17) | 0.500 (6) |
| 64 | 1.525 | 16.3 (9) | 27.7 (20) | 0.500 (6) |
| 64 | 1.550 | 17.4 (10) | 26.1 (18) | 0.500 (6) |
| 64 | 1.575 | 19.6 (12) | 26.9 (19) | 0.500 (6) |
| 64 | 1.600 | 17.6 (10) | 25.3 (17) | 0.500 (6) |
| 128 | 1.300 | 26.4 (18) | 27.5 (19) | 0.500 (6) |
| 128 | 1.400 | 17.5 (10) | 19.9 (12) | 0.500 (6) |
| 128 | 1.500 | 15.1 (8) | 29.5 (22) | 0.500 (6) |
| 128 | 1.525 | 15.6 (8) | 30.1 (22) | 0.500 (6) |
| 128 | 1.550 | 16.9 (9) | 31.4 (24) | 0.500 (6) |
| 128 | 1.575 | 17.0 (9) | 31.8 (24) | 0.500 (5) |
| 128 | 1.600 | 20.9 (13) | 34.6 (28) | 0.500 (5) |
| 128 | 1.625 | 22.0 (14) | 39.2 (28) | 0.500 (5) |
| 128 | 1.650 | 22.0 (14) | 33.2 (26) | 0.500 (6) |
| 128 | 1.675 | 21.3 (13) | 32.4 (25) | 0.500 (6) |
| 128 | 1.700 | 19.5 (12) | 28.6 (21) | 0.500 (5) |
| 128 | 1.725 | 20.7 (13) | 29.7 (22) | 0.500 (5) |
| 256 | 1.300 | 73.1 (85) | 74.4 (87) | 0.500 (6) |
| 256 | 1.400 | 33.7 (26) | 24.6 (16) | 0.503 (6) |
| 256 | 1.500 | 19.6 (12) | 19.6 (12) | 0.500 (6) |
| 256 | 1.600 | 16.6 (9) | 30.3 (23) | 0.500 (6) |
| 256 | 1.625 | 17.9 (10) | 37.2 (31) | 0.500 (6) |
| 256 | 1.650 | 20.1 (12) | 37.4 (31) | 0.500 (6) |
| 256 | 1.675 | 22.2 (14) | 40.8 (35) | 0.500 (6) |
| 256 | 1.700 | 24.1 (16) | 38.0 (32) | 0.500 (6) |
| 256 | 1.725 | 25.3 (17) | 44.5 (40) | 0.500 (6) |
| 256 | 1.750 | 30.1 (22) | 49.2 (47) | 0.500 (6) |
| 256 | 1.775 | 27.4 (19) | 39.6 (34) | 0.500 (5) |
| 256 | 1.800 | 27.4 (19) | 43.1 (38) | 0.500 (6) |
| 256 | 1.825 | 24.4 (16) | 36.7 (30) | 0.500 (6) |
| 256 | 1.850 | 29.2 (21) | 37.3 (31) | 0.500 (6) |
| 512 | 1.400 | 89.9 (115) | 73.8 (86) | 0.500 (6) |
| 512 | 1.500 | 38.5 (32) | 24.7 (17) | 0.500 (6) |
| 512 | 1.600 | 20.3 (12) | 22.8 (15) | 0.500 (5) |
| 512 | 1.650 | 19.1 (11) | 23.3 (15) | 0.500 (6) |
| 512 | 1.700 | 19.1 (11) | 29.6 (22) | 0.500 (6) |
| 512 | 1.750 | 24.3 (16) | 51.1 (49) | 0.500 (5) |
| 512 | 1.800 | 33.7 (26) | 53.5 (53) | 0.500 (5) |
| 1024 | 1.550 | 52.5 (52) | 46.6 (43) | 0.500 (6) |
| 1024 | 1.600 | 32.1 (25) | 25.0 (17) | 0.500 (6) |
| 1024 | 1.700 | 21.2 (13) | 21.0 (13) | 0.500 (6) |
| 1024 | 1.800 | 25.7 (18) | 46.1 (42) | 0.500 (5) |
| 1024 | 1.900 | 41.7 (36) | 78.2 (94) | 0.500 (5) |

TABLE II. Results of numerical simulations for the autocorrelation times with over heat bath algorithm.

| L | $\beta$ | $\tau^{int}_\chi$ | $\tau^{int}_{\chi'}$ | $\tau^{int}_{E_1}$ |
|---|---|---|---|---|
| 32 | 1.250 | 8.8 (4) | 5.8 (2) | 2.37 (6) |
| 32 | 1.300 | 10.9 (6) | 6.3 (2) | 2.70 (7) |
| 32 | 1.350 | 12.4 (7) | 7.4 (3) | 2.75 (7) |
| 32 | 1.400 | 11.2 (6) | 7.3 (3) | 2.48 (6) |
| 32 | 1.450 | 9.8 (5) | 7.6 (3) | 2.11 (5) |
| 32 | 1.500 | 8.8 (4) | 6.9 (3) | 1.93 (4) |
| 64 | 1.200 | 7.8 (3) | 5.9 (2) | 2.16 (5) |
| 64 | 1.250 | 11.1 (6) | 7.7 (3) | 2.33 (6) |
| 64 | 1.300 | 17.7 (12) | 10.1 (5) | 2.38 (6) |
| 64 | 1.350 | 24.8 (19) | 14.7 (9) | 2.62 (7) |
| 64 | 1.400 | 33.4 (30) | 18.7 (13) | 2.60 (7) |
| 64 | 1.450 | 37.8 (36) | 23.6 (18) | 2.58 (7) |
| 64 | 1.500 | 40.4 (40) | 20.7 (15) | 2.32 (6) |
| 64 | 1.525 | 34.9 (32) | 23.5 (18) | 2.28 (5) |
| 64 | 1.550 | 30.1 (26) | 22.1 (16) | 2.14 (5) |
| 64 | 1.575 | 26.0 (21) | 20.9 (15) | 2.04 (5) |
| 64 | 1.600 | 28.2 (23) | 23.1 (17) | 1.95 (4) |
| 128 | 1.200 | 7.0 (3) | 6.8 (3) | 2.38 (6) |
| 128 | 1.300 | 11.3 (6) | 10.8 (6) | 2.45 (6) |
| 128 | 1.400 | 16.8 (10) | 14.5 (9) | 2.63 (7) |
| 128 | 1.500 | 48.7 (53) | 37.6 (36) | 2.66 (7) |
| 128 | 1.525 | 93.0 (139) | 72.9 (97) | 2.59 (7) |
| 128 | 1.550 | 127.6 (223) | 74.4 (100) | 2.47 (6) |
| 128 | 1.575 | 143.1 (265) | 96.4 (147) | 2.24 (5) |
| 128 | 1.600 | 140.1 (257) | 83.0 (117) | 2.26 (5) |
| 128 | 1.625 | 144.3 (269) | 101.1 (158) | 2.22 (5) |
| 128 | 1.650 | 126.6 (221) | 82.4 (116) | 2.03 (4) |
| 128 | 1.675 | 117.3 (197) | 88.6 (129) | 1.98 (4) |
| 128 | 1.700 | 121.3 (207) | 108.0 (174) | 1.92 (4) |
| 128 | 1.725 | 75.0 (101) | 65.6 (82) | 1.91 (4) |
| 256 | 1.300 | 19.1 (13) | 19.1 (13) | 2.45 (6) |
| 256 | 1.400 | 62.2 (76) | 47.8 (51) | 2.60 (7) |
| 256 | 1.500 | 184.7 (275) | 92.3 (97) | 2.58 (5) |
| 256 | 1.600 | 255.5 (447) | 183.6 (272) | 2.33 (4) |
| 256 | 1.625 | 488.9 (1184) | 195.4 (299) | 2.27 (4) |
| 256 | 1.650 | 455.7 (1066) | 319.2 (625) | 2.08 (3) |
| 256 | 1.675 | 582.6 (1541) | 412.6 (918) | 2.06 (3) |
| 256 | 1.700 | 576.9 (2147) | 313.1 (859) | 1.96 (3) |
| 256 | 1.725 | 373.6 (1119) | 247.2 (602) | 1.96 (3) |
| 256 | 1.775 | 377.4 (1136) | 378.5 (1141) | 1.88 (4) |
| 256 | 1.800 | 524.2 (1860) | 246.1 (593) | 1.76 (4) |

TABLE III. Critical slowing down exponents for the generalized algorithm.

| $L_1$ | $L_2$ | $z_{M^2}$ | $z_F$ | $z_{G_1}$ |
|---|---|---|---|---|
| 32 | 64 | 0.19 | 0.11 | $\sim 0.0$ |
| 32 | 128 | 0.24 | 0.23 | $\sim 0.0$ |
| 32 | 256 | 0.26 | 0.25 | $\sim 0.0$ |
| 32 | 512 | 0.27 | 0.23 | $\sim 0.0$ |
| 32 | 1024 | 0.32 | 0.29 | $\sim 0.0$ |
| 64 | 128 | 0.23 | 0.30 | $\sim 0.0$ |
| 64 | 256 | 0.24 | 0.22 | $\sim 0.0$ |
| 64 | 512 | 0.23 | 0.18 | $\sim 0.0$ |
| 64 | 1024 | 0.30 | 0.26 | $\sim 0.0$ |
| 128 | 256 | 0.25 | 0.16 | $\sim 0.0$ |
| 128 | 512 | 0.20 | 0.10 | $\sim 0.0$ |
| 128 | 1024 | 0.31 | 0.21 | $\sim 0.0$ |
| 256 | 512 | 0.24 | 0.16 | $\sim 0.0$ |
| 256 | 1024 | 0.35 | 0.20 | $\sim 0.0$ |
| 512 | 1024 | 0.47 | 0.38 | $\sim 0.0$ |

TABLE IV. Critical slowing down exponents for the over heat bath algorithm.

| $L_1$ | $L_2$ | $z_{M^2}$ | $z_F$ | $z_{G_1}$ |
|---|---|---|---|---|
| 32 | 64 | 1.65 | 1.63 | $\sim 0.0$ |
| 32 | 128 | 1.65 | 1.73 | $\sim 0.0$ |
| 32 | 256 | 1.69 | 1.76 | $\sim 0.0$ |
| 64 | 128 | 1.81 | 1.91 | $\sim 0.0$ |
| 64 | 256 | 1.43 | 1.83 | $\sim 0.1$ |
| 128 | 256 | 1.66 | 1.67 | $\sim 0.2$ |

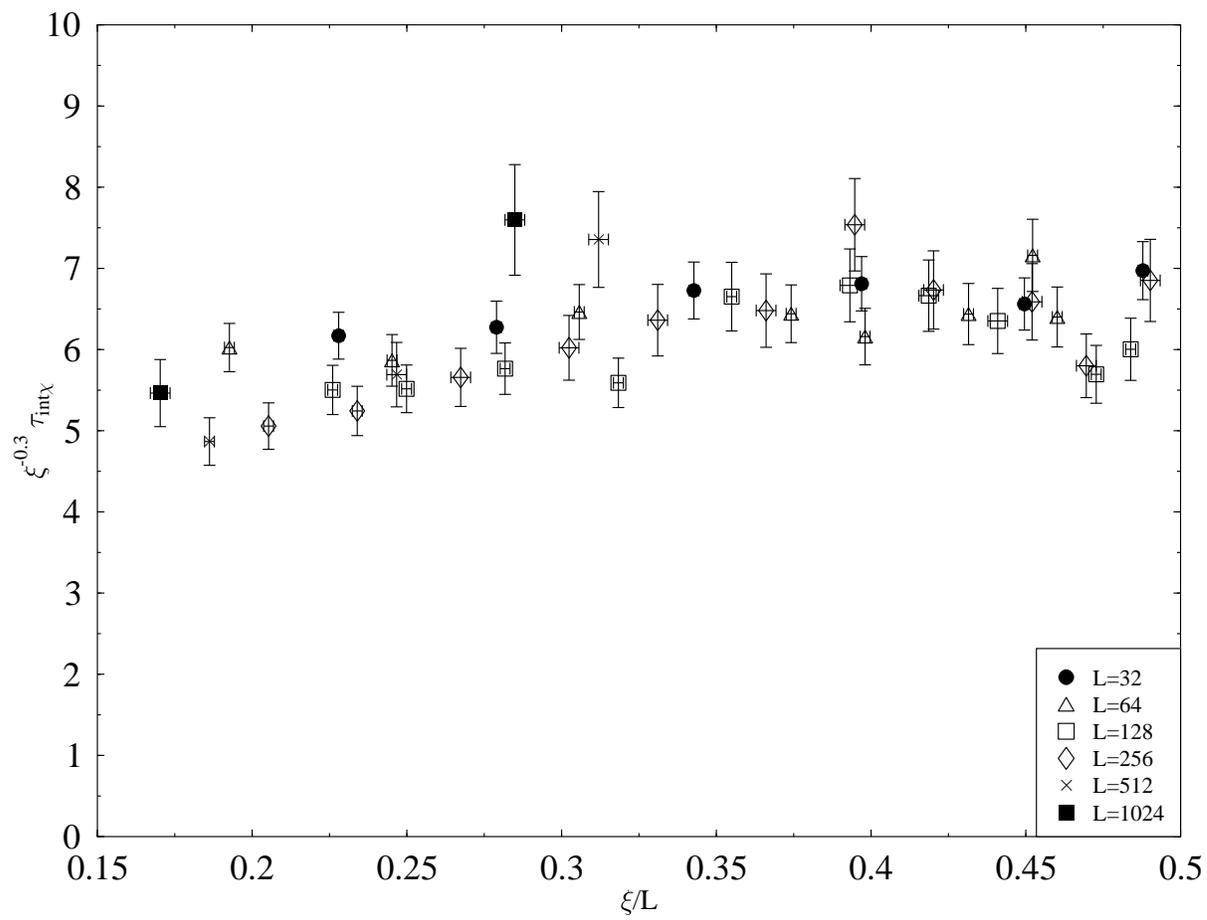

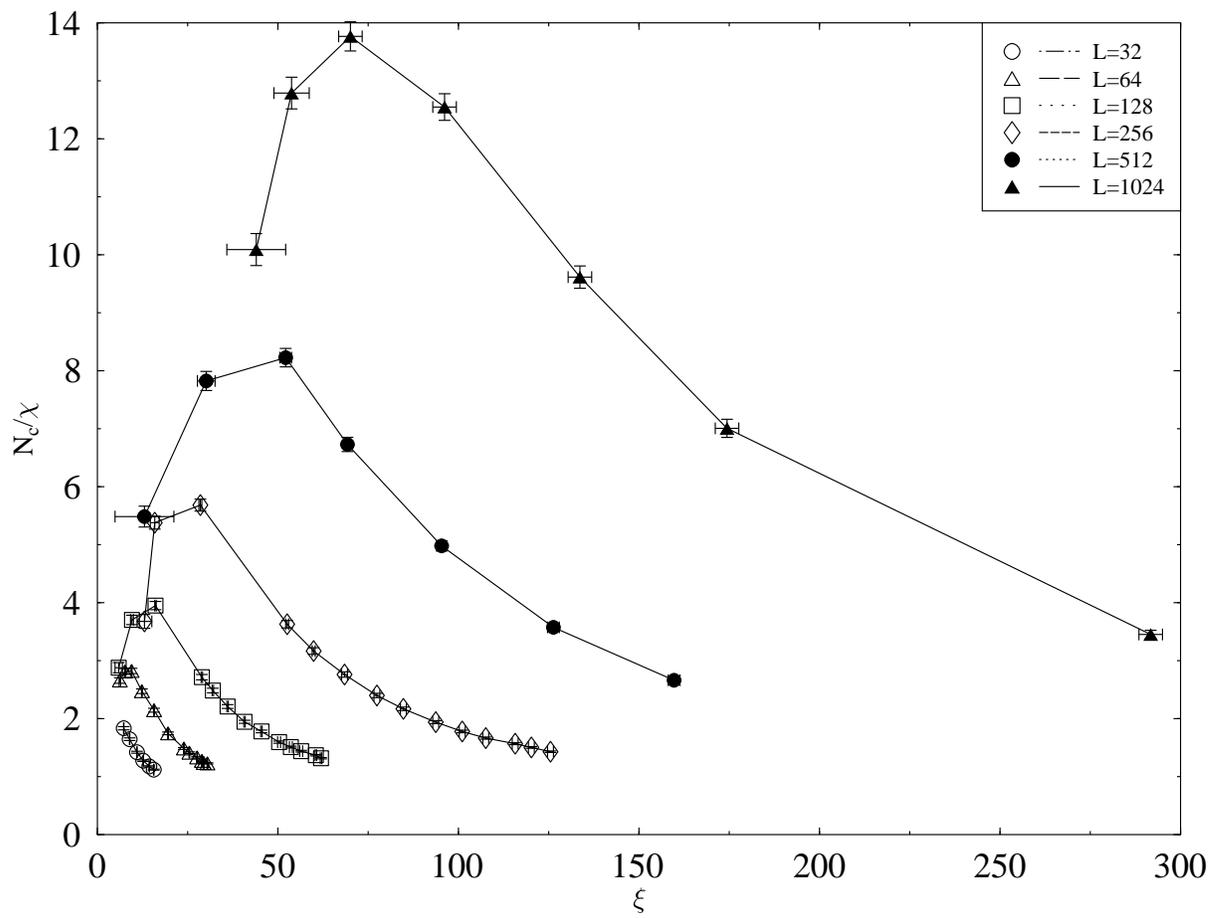